\documentclass[11pt]{article}
\usepackage{latexsym,amsmath,amsfonts,graphics}
\begin{document}
\begin{center}
{\LARGE\textbf{Synchronization in Networks of Coupled Harmonic Oscillators with Stochastic Perturbation and Time Delays}}\\
\bigskip
\bigskip
Yilun Shang\footnote{Department of Mathematics, Shanghai Jiao Tong University, Shanghai 200240, CHINA. email: \texttt{shyl@sjtu.edu.cn}}\\
\end{center}

\begin{abstract}
In this paper, we investigate synchronization of coupled
second-order linear harmonic oscillators with random noises and time
delays. The interaction topology is modeled by a weighted directed
graph and the weights are perturbed by white noise. On the basis of
stability theory of stochastic differential delay equations,
algebraic graph theory and matrix theory, we show that the coupled
harmonic oscillators can be synchronized almost surely with
perturbation and time delays. Numerical examples are presented to
illustrate our theoretical results.

\bigskip

\end{abstract}

\bigskip
\normalsize

\noindent{\Large\textbf{1. Introduction}}
\smallskip

Synchronization, as an emergent collective phenomenon of a
population of units with oscillatory behaviors, is one of the most
intriguing in nature and plays a significant role in a variety of
disciplines such as biology, sociology, physics, chemistry and
technology \cite{7,9,8}. One celebrated model for synchronization is
the Kuramoto model \cite{10}, which is mathematically tractable and
described by a system of structured ordinary differential equations.
The original Kuramoto formulation assumes full connectivity of the
network, i.e. whose interaction topology is a complete graph. Recent
works generalize the Kuramoto model to nearest neighbor interaction
and the underlying topologies may be general complex networks, see
e.g. \cite{12,14,13}.

Another classical model for synchronization is the harmonic
oscillator network \cite{15,1,11}, which is the subject of this
paper. Recently, Ren \cite{1} investigates synchronization of
coupled second-order linear harmonic oscillators with local
interaction. Due to the linear structure, the ultimate trajectories
to which each oscillator converges over directed fixed networks are
shown explicitly and milder convergence conditions than those in the
case of Kuramoto model can be derived. Consensus problems, known as
the ability of an ensemble of dynamic agents to reach a common value
in an asymptotical way, are prominent applications of
synchronization in engineering and computer science; see \cite{2}
for a recent survey. The harmonic oscillator networks are related to
the second-order consensus protocols addressed in
\cite{17,16,18,19}. Compared with these works, where the consensus
equilibrium for the velocities of agents is a constant, the
positions and velocities are synchronized to achieve oscillating
motion by utilizing harmonic oscillator schemes (c.f. Remark 4
below).

Since noise is ubiquitous in nature,technology, and society, the
motion of oscillator is inevitably subject to disturbance in the
environments. On the other hand, time delay is also unavoidable
because the information spreading through a complex network is
characterized by the finite speeds of signal transmission over a
distance. Although random noise and time delay have been studied
extensively in exploring synchronization (or consensus problems) by
means of theoretical and numerical methods, they have seldom been
analytically treated in synchronization of coupled harmonic
oscillators. Motivating this idea, the objective of this paper is to
deal with synchronization conditions for coupled harmonic
oscillators over general directed topologies with noise perturbation
and communication time delays. The main tools used here are
algebraic graph theory, matrix theory and stochastic differential
delay equation theory.

The rest of the paper is organized as follows. In Section 2, we
provide some preliminaries and present the coupled harmonic
oscillator network model. In Section 3, we analyze the
synchronization stability of this model and give sufficient
conditions for almost surely convergence. Numerical examples are
given in Section 4 to validate our theoretical results. Finally, the
conclusion is drawn in Section 5.

\bigskip
\noindent{\Large\textbf{2. Problem formulation}}
\smallskip

By convention, $\mathbb{R}$ represents the real number set; $I_n$ is
an $n\times n$ identity matrix. For any vector $x$, $x^T$ denotes
its transpose and its norm $\|x\|$ is the Euclidean norm. For a
matrix $A$, denote by $\|A\|$ the operator norm of $A$, i.e.
$\|A\|=\sup\{\|Ax\|:\|x\|=1\}$. $\mathrm{Re}(z)$ denotes the real
part of $z\in\mathbb{C}$.

Throughout the paper we will be using the following concepts on
graph theory (see e.g. \cite{4}) to capture the topology of the
network interactions.

Let $\mathcal{G}=(\mathcal{V},\mathcal{E},\mathcal{A})$ be a
weighted directed graph with the set of vertices
$\mathcal{V}=\{1,2,\cdots,n\}$ and the set of arcs
$\mathcal{E}\subseteq\mathcal{V}\times\mathcal{V}$. The vertex $i$
in $\mathcal{G}$ represents the $i$th oscillator, and a directed
edge $(i,j)\in\mathcal{E}$ means that oscillator $j$ can directly
receive information from oscillator $i$. The set of neighbors of
vertex $i$ is denoted by $\mathcal{N}_i=\{j\in\mathcal{V}|\
(j,i)\in\mathcal{E}\}$. $\mathcal{A}=(a_{ij})\in\mathbb{R}^{n\times
n}$ is called the weighted adjacency matrix of $\mathcal{G}$ with
nonnegative elements and $a_{ij}>0$ if and only if
$j\in\mathcal{N}_i$. The in-degree of vertex $i$ is defined as
$d_i=\sum_{j=1}^na_{ij}$. The Laplacian of $\mathcal{G}$ is defined
as $L=D-\mathcal{A}$, where $D=\mathrm{diag}(d_1, d_2,\cdots, d_n)$.
A directed graph $\mathcal{G}$ is called strongly connected if there
is a directed path from $i$ to $j$ between any two distinct vertices
$i,j\in\mathcal{V}$. There exists a directed path from vertex $i$ to
vertex $j$, then $j$ is said to be reachable from $i$. For every
vertex in directed graph $\mathcal{G}$, if there is a path from
vertex $i$ to it, then we say $i$ is globally reachable in
$\mathcal{G}$. In this case, we also say that $\mathcal{G}$ has a
directed spanning tree with root $i$.

Consider $n$ coupled harmonic oscillators connected by dampers and
each attached to fixed supports by identical springs with spring
constant $k$. The resultant dynamical system can be described as
\begin{equation}
\ddot{x_i}+kx_i+\sum_{j\in\mathcal{N}_i}a_{ij}\big(\dot{x_i}-\dot{x_j}\big)=0,\quad
i=1,\cdots,n\label{1}
\end{equation}
where $x_i\in\mathbb{R}$ denotes the position of the $i$th
oscillator, $k$ serves as a positive gain, and $a_{ij}$
characterizes interaction between oscillators $i$ and $j$ as
mentioned before.

Here we study a leader-follower version (c.f. Remark 1) of the above
system, and moreover, communication time delay and stochastic noises
during the propagation of information from oscillator to oscillator
are introduced. In particular, we consider the dynamical system of
the form:

\begin{multline}
\ddot{x_i}(t)+kx_i(t)+\sum_{j\in\mathcal{N}_i}a_{ij}\big(\dot{x_i}(t-\tau)-\dot{x_j}(t-\tau)\big)
+b_i\big(\dot{x_i}(t-\tau)-\dot{x_0}(t-\tau)\big)\\
+\big[\sum_{j\in\mathcal{N}_i}\sigma_{ij}\big(\dot{x_i}(t-\tau)-\dot{x_j}(t-\tau)\big)
+\rho_i\big(\dot{x_i}(t-\tau)-\dot{x_0}(t-\tau)\big)\big]\dot{w_i}(t)=0,\quad
i=1,\cdots,n,\label{2}
\end{multline}

\begin{equation}
\ddot{x_0}(t)+kx_0(t)=0,\label{3}
\end{equation}
where $\tau$ is the time delay and $x_0$ is the position of the
virtual leader, labeled as oscillator 0, which follows Equation
(\ref{3}) describing an undamped harmonic oscillator. We thus
concern another directed graph
$\overline{\mathcal{G}}\supset\mathcal{G}$ associated with the
system consisting of $n$ oscillators and one leader. Let
$B=\mathrm{diag}(b_1,\cdots,b_n)$ be a diagonal matrix with
nonnegative diagonal elements and $b_i>0$ if and only if
$0\in\mathcal{N}_i$. Let $W(t):=(w_1(t),\cdots,w_n(t))^T$ be an
$n$-dimensional standard Brownian motion. Hence, $\dot{w_i}(t)$ is
one-dimensional white noise. To highlight the presence of noise, it
is natural to assume that $\sigma_{ij}>0$ if $j\in\mathcal{N}_i$,
and $\sigma_{ij}=0$ otherwise; $\rho_{ij}>0$ if $j\in\mathcal{N}_i$,
and $\rho_{ij}=0$ otherwise. Also let
$A_{\sigma}=(\sigma_{ij})\in\mathbb{R}^{n\times n}$ and
$B_{\sigma}=\mathrm{diag}(\rho_1,\cdots,\rho_n)$ be two matrices
representing the intensity of noise. Moreover, let
$\sigma_i=\sum_{j=1}^n\sigma_{ij}$,
$D_{\sigma}=\mathrm{diag}(\sigma_1,\cdots,\sigma_n)$, and
$L_{\sigma}=D_{\sigma}-A_{\sigma}$.

\smallskip
\noindent\textbf{Remark 1.}\itshape \quad Consensus problems of
self-organized groups with leaders have broad applications in
swarms, formation control and robotic systems, etc.; see e.g.
\cite{20,21,22}. In multi-agent systems, the leaders have influence
on the followers' behaviors but usually independent of their
followers. One therefore transfers the control of a whole system to
that of a single agent, which saves energy and simplifies network
control design \cite{23,24}. Most of the existing relevant
literatures assume a constant state leader, while our model serves
to be an example of oscillating state leader on this stage.
 \normalfont

\smallskip
Let $r_i=x_i$ and $v_i=\dot{x_i}$ for $i=0,1,\cdots,n$. By denoting
$r=(r_1,\cdots,r_n)^T$ and $v=(v_1,\cdots,v_n)^T$, we can rewrite
the system (\ref{2}), (\ref{3}) in a compact form as:
\begin{eqnarray}
\mathrm{d}r(t)&=&v(t)\mathrm{d}t,\label{4}\\
\mathrm{d}v(t)&=&\big[-kr(t)-(L+B)v(t-\tau)+Bv_0(t-\tau)1\big]\mathrm{d}t\nonumber\\
&
&+\big[-(L_{\sigma}+B_{\sigma})v(t-\tau)+B_{\sigma}v_0(t-\tau)1\big]\mathrm{d}W,\label{5}\\
\mathrm{d}r_0(t)&=&v_0(t)\mathrm{d}t,\quad
\mathrm{d}v_0(t)=-kr_0(t)\mathrm{d}t,\label{6}
\end{eqnarray}
where $1$ denotes an $n\times1$ column vector of all ones (with some
ambiguity; however, the right meaning would be clear in the
context).

\smallskip
\noindent\textbf{Remark 2.}\itshape \quad Note that $v_i$ depends on
the information from its in-neighbors and itself. In the special
case that time delay $\tau=0$ and $A_{\sigma}=B_{\sigma}=0$,
algorithms (\ref{4})-(\ref{6}) are equivalent to algorithms (12) and
(13) in \cite{1}. \normalfont

\bigskip
\noindent{\Large\textbf{3. Convergence analysis}}
\smallskip

In this section, the convergence analysis of systems
(\ref{4})-(\ref{6}) is given and we show that $n$ coupled harmonic
oscillators (followers) are synchronized to the oscillating behavior
of the virtual leader with probability one.

Before proceeding, we introduce an exponential stability result for
the following $n$-dimensional stochastic differential delay equation
(for more details, see e.g. \cite{6})

\begin{equation}
\mathrm{d}x(t)=[Ex(t)+Fx(t-\tau)]\mathrm{d}t+g(t,x(t),x(t-\tau))\mathrm{d}W(t),\label{7}
\end{equation}
where $E$ and $F$ are $n\times n$ matrices, $g:[0,\infty)\times
\mathbb{R}^n\times\mathbb{R}^n\rightarrow\mathbb{R}^{n\times m}$
which is locally Lipschitz continuous and satisfies the linear
growth condition with $g(t,0,0)\equiv0$, $W(t)$ is an
$m$-dimensional standard Brownian motion.

\smallskip
\noindent\textbf{Lemma 1.}(\cite{5})\itshape \quad Assume that there
exists a pair of symmetric positive definite $n\times n$ matrices
$P$ and $Q$ such that $P(E+F)+(E+F)^TP=-Q$. Assume also that there
exist non-negative constants $\alpha$ and $\beta$ such that
\begin{equation}
\mathrm{trace}[g^T(t,x,y)g(t,x,y)]\le\alpha\|x\|^2+\beta\|y\|^2\label{8}
\end{equation}
for all $(t,x,y)\in[0,\infty)\times \mathbb{R}^n\times\mathbb{R}^n$.
Let $\lambda_{\min}(Q)$ be the smallest eigenvalue of $Q$. If
$$
(\alpha+\beta)\|P\|+2\|PF\|\sqrt{2\tau(4\tau(\|E\|^2+\|F\|^2)+\alpha+\beta)}<\lambda_{\min}(Q),
$$
then the trivial solution of Equation (\ref{7}) is almost surely
exponentially stable.
\normalfont
\smallskip

We need the following lemma for Laplacian matrix.

\smallskip
\noindent\textbf{Lemma 2.}(\cite{3})\itshape \quad Let $L$ be the
Laplacian matrix associated with a directed graph $\mathcal{G}$.
Then $L$ has a simple zero eigenvalue and all its other eigenvalues
have positive real parts if and only if $\mathcal{G}$ has a directed
spanning tree. In addition, $L1=0$ and there exists
$p\in\mathbb{R}^n$ satisfying $p\ge0$, $p^TL=0$ and $p^T1=1$.
\normalfont
\smallskip

Let
$$
\left\{
\begin{array}{cc}
r_0(t):=&\cos(\sqrt{k}t)r_0(0)+\frac1{\alpha}\sin(\sqrt{k}t)v_0(0),\\
v_0(t):=&-\sqrt{k}\sin(\sqrt{k}t)r_0(0)+\cos(\sqrt{k}t)v_0(0).
\end{array}
\right.
$$
Then it is easy to see that $r_0(t)$ and $v_0(t)$ solve (\ref{6}).
Let $r^*=r-r_01$, $v^*=v-v_01$. Invoking Lemma 2, we can obtain an
error dynamics of (\ref{4})-(\ref{6}) as follows

\begin{equation}
\mathrm{d}\varepsilon(t)=[E\varepsilon(t)+F\varepsilon(t-\tau)]\mathrm{d}t+H\varepsilon(t-\tau)\mathrm{d}W(t),\label{9}
\end{equation}
where
$$
\varepsilon=\left(\begin{array}{c}r^*\\v^*
\end{array}\right),\quad
E=\left(\begin{array}{cc}0&I_n\\-kI_n&0
\end{array}\right),\quad
F=\left(\begin{array}{cc}0&0\\0&-L-B
\end{array}\right),\quad
H=\left(\begin{array}{cc}0&0\\0&-L_{\sigma}-B_{\sigma}
\end{array}\right)
$$
and $W(t)$ is an $2n$-dimensional standard Brownian motion.

Now we present our main result as follows.

\smallskip
\noindent\textbf{Theorem 1.}\itshape \quad Suppose that vertex 0 is
globally reachable in $\overline{\mathcal{G}}$. If
\begin{equation}
\|H\|^2\|P\|+2\|PF\|\sqrt{8\tau^2[(k\vee1)^2+\|F\|^2]+2\tau\|H\|^2}<\lambda_{\min}(Q),\label{10}
\end{equation}
where $k\vee1:=\max\{k,1\}$, $P$ and $Q$ are two symmetric positive
definite matrices such that $P(E+F)+(E+F)^TP=-Q$, then by using
algorithms (\ref{4})-(\ref{6}), we have
$$r(t)-r_0(t)1\rightarrow0,\quad v(t)-v_0(t)1\rightarrow0$$
almost surely, as $t\rightarrow\infty$. Here, $r_0$ and $v_0$ are
given as above. \normalfont

\medskip
\noindent\textbf{Proof}. Clearly, it suffices to prove the trivial
solution $\varepsilon(t;0)=0$ of (\ref{9}) is almost surely
exponential stable.

Let $\{\lambda_i:i=1,\cdots,n\}$ be the set of eigenvalues of
$-L-B$. Since vertex 0 is globally reachable in
$\overline{\mathcal{G}}$, from Lemma 2 it follows that $-L-B$ is a
stable matrix, that is, $\mathrm{Re}(\lambda_i)<0$ for all $i$.

Let $\mu$ be an eigenvalue of matrix $E+F$ and
$\varphi=(\varphi_1^T,\varphi_2^T)^T$ be an associated eigenvector.
We thus have
$$
\left(\begin{array}{cc}0&I_n\\-kI_n&-L-B\end{array}\right)\left(\begin{array}{c}
\varphi_1\\\varphi_2\end{array}\right)=\mu\left(\begin{array}{c}
\varphi_1\\\varphi_2\end{array}\right),
$$
which yields $(-L-B)\varphi_1=\frac{\mu^2+k}{\mu}\varphi_1$ and
$\varphi_1\not=0$. Hence $\mu$ satisfies $\mu^2-\lambda_i\mu+k=0$.
The $2n$ eigenvalues of $E+F$ are shown to be given by
$\mu_{i\pm}=\frac{\lambda_i\pm\sqrt{\lambda_i^2-4k}}{2}$ for
$i=1,\cdots,n$. Since $\mathrm{Re}(\lambda_i)<0$, we get
$\mathrm{Re}(\mu_{i-})=\mathrm{Re}\big(\frac{\lambda_i-\sqrt{\lambda_i^2-4k}}{2}\big)<0$
for $i=1,\cdots,n$. From $\mu_{i+}\mu_{i-}=k$ it follows that
$\mu_{i+}$ and $\mu_{i-}$ are symmetric with respect to the real
axis in the complex plane. Accordingly, $\mathrm{Re}(\mu_{i+})<0$
for $i=1,\cdots,n$; furthermore, $E+F$ is a stable matrix. By
Lyapunov theorem, for all symmetric positive definite matrix $Q$
there exists a unique symmetric positive definite matrix $P$ such
that
\begin{equation}
P(E+F)+(E+F)^TP=-Q.\label{11}
\end{equation}

On the other hand, we have
$\mathrm{trace}(\varepsilon^TH^TH\varepsilon)\le\|H\|^2\|\varepsilon\|^2$.
Therefore, (\ref{8}) holds with $\alpha=0$ and $\beta=\|H\|^2$. Note
that $\|E\|=k\vee1$. We then complete our proof by employing Lemma
1. $\Box$

\smallskip
\noindent\textbf{Remark 3.}\itshape \quad Note that the result of
Theorem 1 is dependent of the choice of matrices $P$ and $Q$. From
computational points of view, the solution to Lyapunov matrix
equation (\ref{11}) may be expressed by using Kronecker product;
$\|H\|=\|L_{\sigma}+B_{\sigma}\|$ and $\|F\|=\|L+B\|$ hold.
\normalfont

\smallskip
\noindent\textbf{Remark 4.}\itshape \quad The algorithms
(\ref{4})-(\ref{6}) can also be applied to synchronized motion
coordination of multi-agent systems, as indicated in \cite{1}
Section 5.
 \normalfont

\smallskip
When deviations between oscillator states exist, we may exploit the
following algorithm to take the place of Equation (\ref{5}):
\begin{eqnarray}
\mathrm{d}v(t)&=&\big[-k(r(t)-\delta)-(L+B)v(t-\tau)+Bv_0(t-\tau)1\big]\mathrm{d}t\nonumber\\
&
&+\big[-(L_{\sigma}+B_{\sigma})v(t-\tau)+B_{\sigma}v_0(t-\tau)1\big]\mathrm{d}W,\label{12}
\end{eqnarray}
where $\delta=(\delta_1,\cdots,\delta_n)^T$ is a constant vector
denoting the deviations. Similarly, we obtain the following result.

\smallskip
\noindent\textbf{Corollary 1.}\itshape \quad Suppose that vertex 0
is globally reachable in $\overline{\mathcal{G}}$, and condition
(\ref{10}) holds, then by using algorithms (\ref{4}), (\ref{6}) and
(\ref{12}), we have
$$r(t)-\delta-r_0(t)1\rightarrow0,\quad v(t)-v_0(t)1\rightarrow0$$
almost surely, as $t\rightarrow\infty$. Here, $r_0$ and $v_0$ are
defined as in Theorem 1. \normalfont

\bigskip
\noindent{\Large\textbf{4. Numerical examples}}
\smallskip

In this section, we provide numerical simulations to illustrate our
results.

We consider a network $\overline{\mathcal{G}}$ consisting of five
coupled harmonic oscillators including one leader indexed by 0 and
four followers as shown in Fig. 1. We assume that $a_{ij}=1$ if
$j\in\mathcal{N}_i$ and $a_{ij}=0$ otherwise; $b_i=1$ if
$0\in\mathcal{N}_i$ and $b_i=0$ otherwise. Note that vertex 0 is
globally reachable in $\overline{\mathcal{G}}$. For simplicity, we
take the noise intensity matrices $L_{\sigma}=0.1L$ and
$B_{\sigma}=0.1B$. We take $Q=I_8$ with $\lambda_{\min}(Q)=1$. By
straightforward calculation, it is obtained that $\|H\|=0.2466$ and
$\|F\|=2.4656$. Two different gains $k$ are explored as follows:

Firstly, we take $k=0.6$ such that $\|E\|=1>k$. We solve $P$ from
Equation (\ref{11}) and get $\|P\|=8.0944$ and $\|PF\|=4.1688$.
Hence the condition (\ref{10}) in Theorem 1 is satisfied by taking
time delay $\tau=0.002$. Thus, the oscillator states are
synchronized successfully as shown in Fig. 2 and Fig. 3 with initial
values given by $\varepsilon(0)=(-5,1,4,-3,-8,2,-1.5,3)^T$.

Secondly, we take $k=2$ such that $\|E\|=k>1$. In this case we
obtain $\|P\|=8.3720$, $\|PF\|=7.5996$ and the condition (\ref{10})
is satisfied by taking time delay $\tau=0.001$. Thereby the
oscillator states are synchronized successfully as shown in Fig. 4
and Fig. 5 with the same initial values given as above.

We see that the value of $k$ not only has an effect on the magnitude
and frequency of the synchronized states (as implied in Theorem 1),
but also affects the shapes of synchronization error curves
$\|r^*\|$ and $\|v^*\|$.

\bigskip
\noindent{\Large\textbf{5. Conclusion}}
\smallskip

This paper is concerned with synchronization of coupled harmonic
oscillators with stochastic perturbation and time delays. Based on
the stability theory of stochastic differential delay equations, we
have shown that the coupled second-order linear harmonic oscillators
are synchronized (i.e. follow the leader) with probability one
provided the leader is globally reachable and the time delay is
sufficient small. Numerical simulations are presented to illustrate
our theoretical results. Since we only investigate the case when the
network topology is fixed, how to consider the time varying topology
is our future research.

\bigskip
\bigskip

\begin{center}
\textbf{Figure captions}

\end{center}

Fig. 1\quad Directed network $\overline{\mathcal{G}}$ for five
coupled harmonic oscillators involving one leader.
$\overline{\mathcal{G}}$ has $0-1$ weights.

Fig. 2\quad Synchronization error $\|r^*\|$ for $k=0.6$ and
$\tau=0.002$.

Fig. 3\quad Synchronization error $\|v^*\|$ for $k=0.6$ and
$\tau=0.002$.

Fig. 4\quad Synchronization error $\|r^*\|$ for $k=2$ and
$\tau=0.001$.

Fig. 5\quad Synchronization error $\|v^*\|$ for $k=2$ and
$\tau=0.001$.

\bigskip
\bigskip

\begin{center}
\setlength{\unitlength}{1mm}
\begin{picture}(60,60)
\put(2,58){0}\put(2,2){1}\put(58,2){3}\put(58,58){2}
\put(2,57){\vector(0,-1){51}}\put(5,59){\vector(1,0){52}}\put(5,5){\vector(1,1){52}}\put(55,55){\vector(-1,-1){50}}\put(5,3){\vector(1,0){52}}\put(59,6){\vector(0,1){51}}
\end{picture}
\end{center}

\centering \scalebox{0.5}{\includegraphics{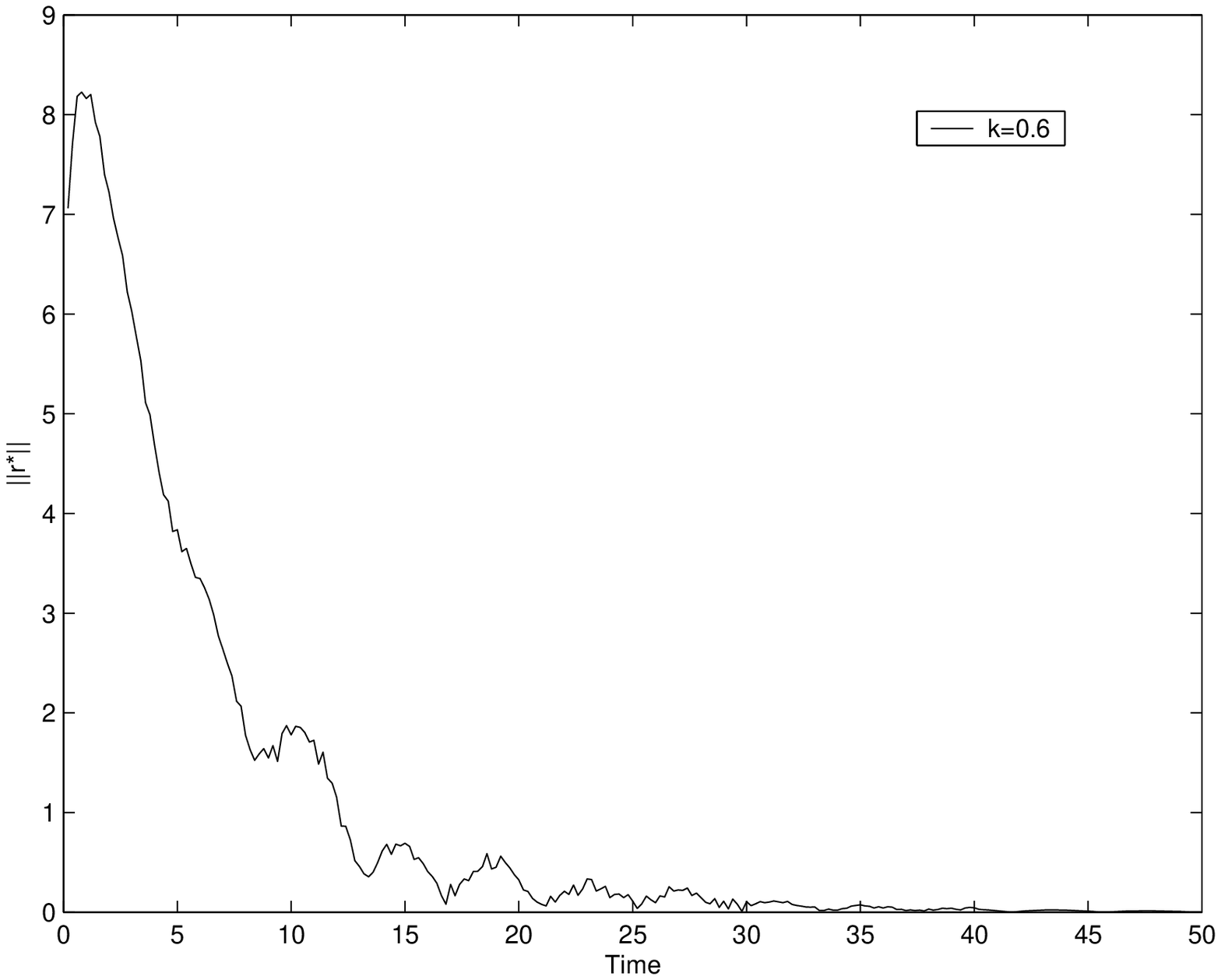}}
 \label{fig_sim} \centering
\scalebox{0.5}{\includegraphics{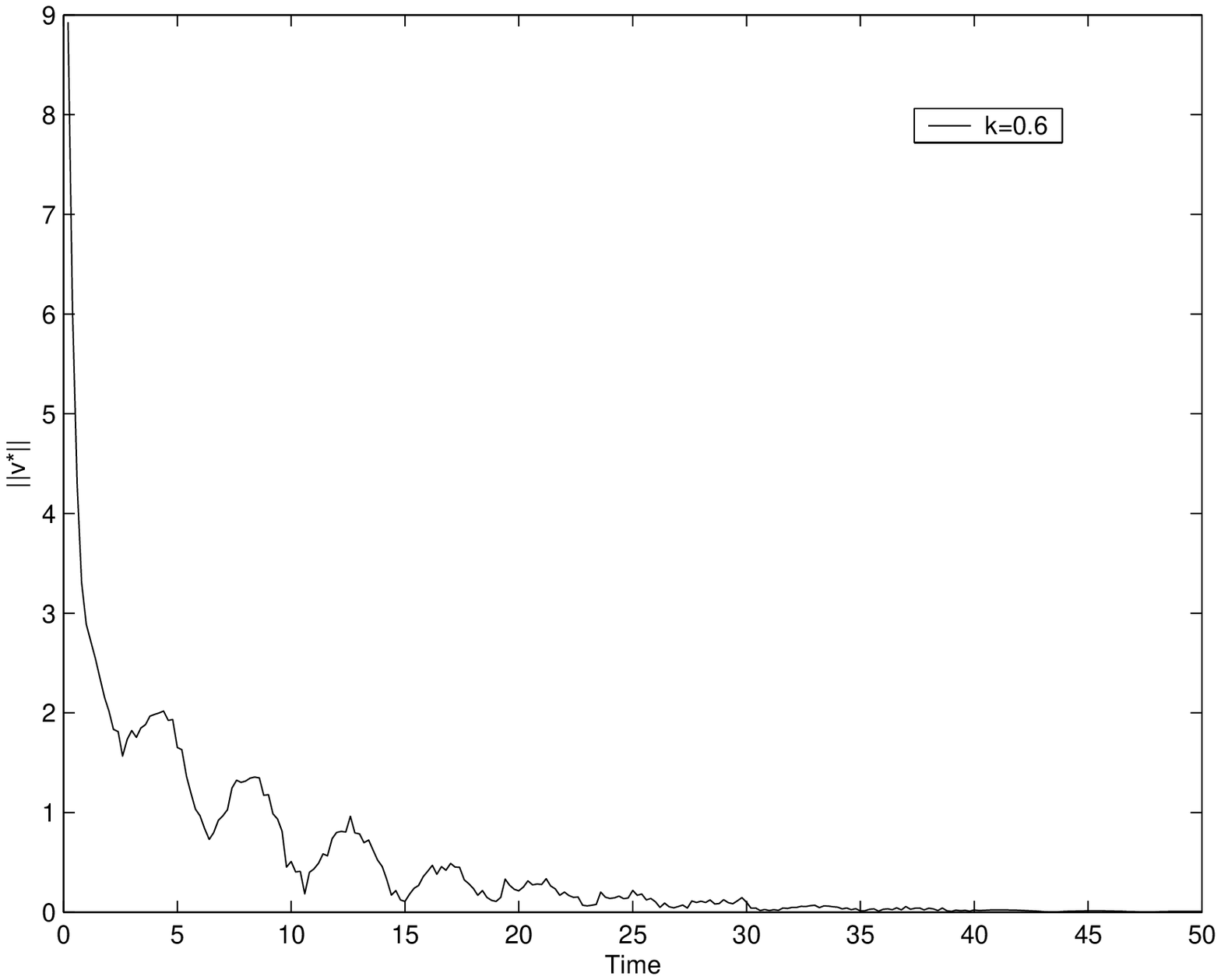}} \label{fig_sim}
\centering \scalebox{0.5}{\includegraphics{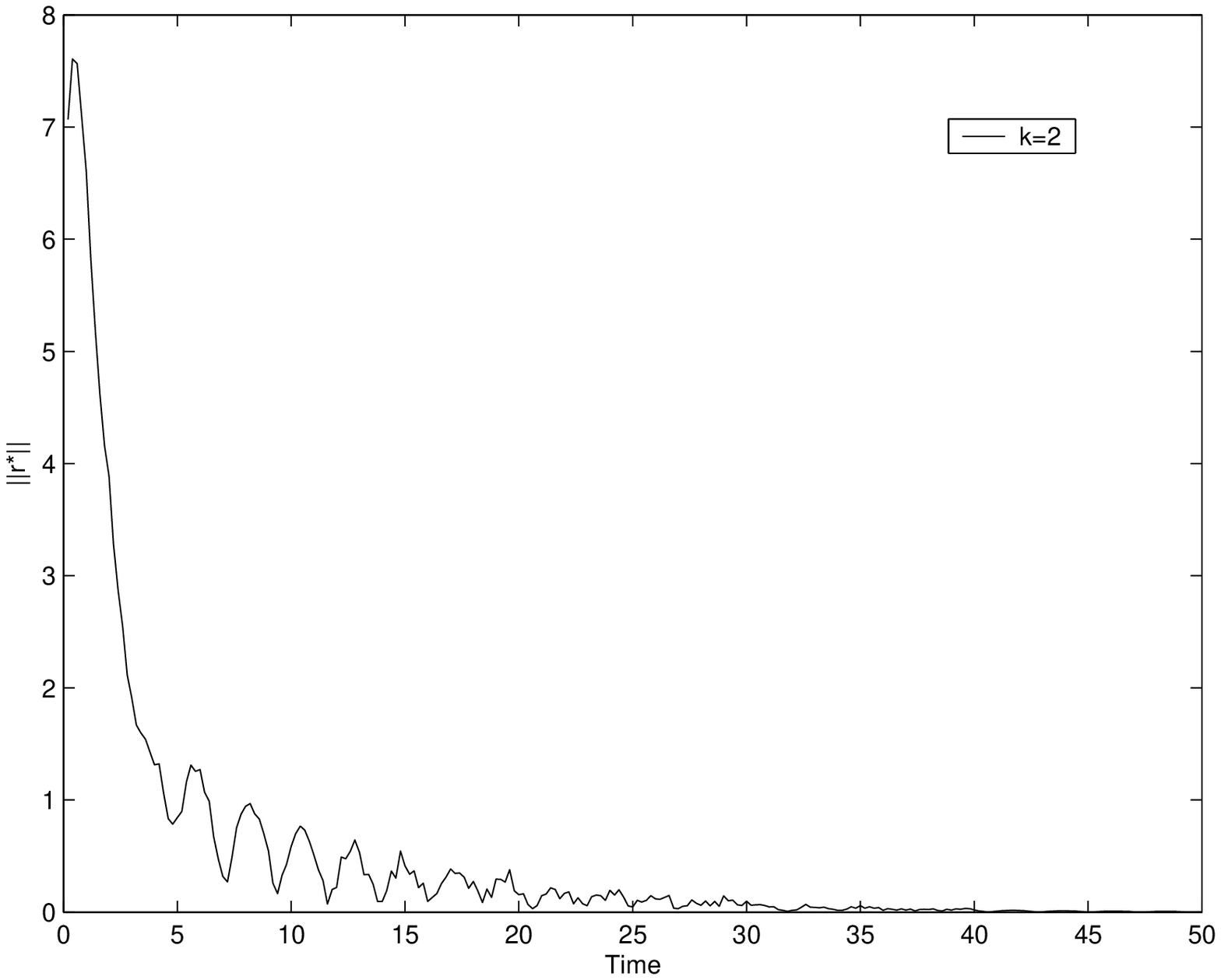}}
 \label{fig_sim} \centering
\scalebox{0.5}{\includegraphics{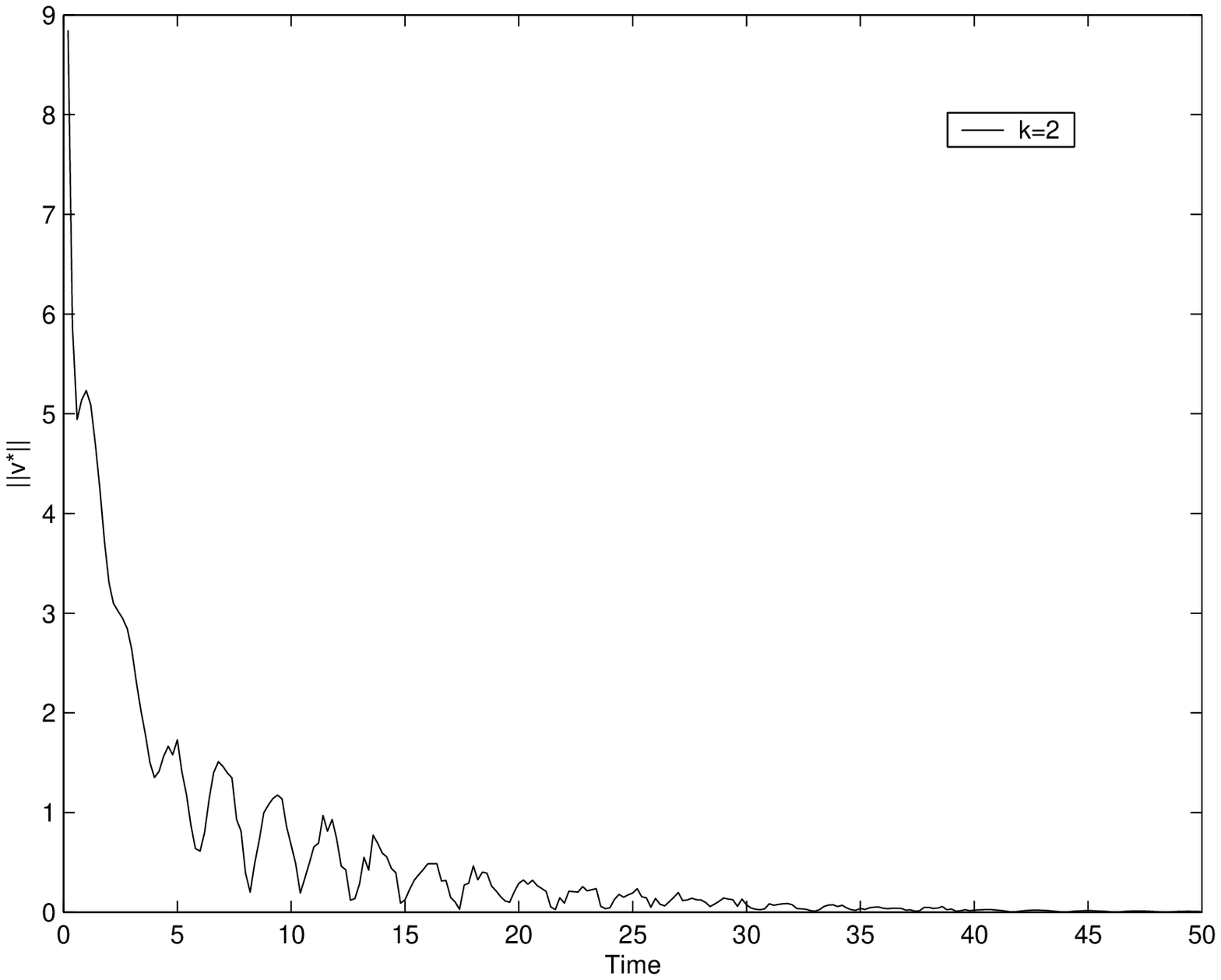}} \label{fig_sim}

\end{document}